\documentclass[%
 reprint,
superscriptaddress,
 amsmath,amssymb,
 aps,
]{revtex4-1}
\usepackage{graphicx,import}
\usepackage{dcolumn}
\usepackage{bm}
\usepackage[rgb]{xcolor}
\usepackage{transparent}
\usepackage[normalem]{ulem}
\usepackage{hyperref}

\usepackage{siunitx}

\begin{document}

\preprint{APS/123-QED}

\title{Fully passive entanglement based quantum key distribution scheme}


\author{Bo Liu}
\email{liubo08@nudt.edu.cn}
\affiliation{College of Advanced Interdisciplinary Studies, National University of Defense Technology, Changsha, 410073, China}
\affiliation{Institute for Quantum Optics and Quantum Information (IQOQI), Austrian Academy of Sciences, Boltzmanngasse 3, 1090 Vienna, Austria}
\affiliation{Vienna Center for Quantum Science and Technology (VCQ), Faculty of Physics, University of Vienna, Boltzmanngasse 5, A-1090 Vienna, Austria}

\author{Matej Pivoluska}
\affiliation{Institute of Computer Science, Masaryk University, Czech Republic
}
\affiliation{Institute of Physics, Slovak Academy of Sciences, Bratislava, Slovakia}

\author{Johannes Handsteiner}
\affiliation{Institute for Quantum Optics and Quantum Information (IQOQI), Austrian Academy of Sciences, Boltzmanngasse 3, 1090 Vienna, Austria}
\affiliation{Vienna Center for Quantum Science and Technology (VCQ), Faculty of Physics, University of Vienna, Boltzmanngasse 5, A-1090 Vienna, Austria}

\author{Dominik Rauch}
\affiliation{Institute for Quantum Optics and Quantum Information (IQOQI), Austrian Academy of Sciences, Boltzmanngasse 3, 1090 Vienna, Austria}
\affiliation{Vienna Center for Quantum Science and Technology (VCQ), Faculty of Physics, University of Vienna, Boltzmanngasse 5, A-1090 Vienna, Austria}


\author{Marcus Huber}
\affiliation{Institute for Quantum Optics and Quantum Information (IQOQI), Austrian Academy of Sciences, Boltzmanngasse 3, 1090 Vienna, Austria}
\affiliation{Vienna Center for Quantum Science and Technology, Atominstitut, Technische Universit$\ddot{a}$t Wien, 1020 Vienna, Austria}

\author{Fabian Steinlechner}
\affiliation{Institute for Quantum Optics and Quantum Information (IQOQI), Austrian Academy of Sciences, Boltzmanngasse 3, 1090 Vienna, Austria}
\affiliation{Vienna Center for Quantum Science and Technology (VCQ), Faculty of Physics, University of Vienna, Boltzmanngasse 5, A-1090 Vienna, Austria}
\affiliation{Fraunhofer Institute for Applied Optics and Precision Engineering IOF, Albert-Einstein-Strasse 7, 07745 Jena, Germany}

\author{Rupert Ursin}
\affiliation{Institute for Quantum Optics and Quantum Information (IQOQI), Austrian Academy of Sciences, Boltzmanngasse 3, 1090 Vienna, Austria}
\affiliation{Vienna Center for Quantum Science and Technology (VCQ), Faculty of Physics, University of Vienna, Boltzmanngasse 5, A-1090 Vienna, Austria}

\author{Thomas Scheidl}
\email{thomas.scheidl@univie.ac.at}
\affiliation{Institute for Quantum Optics and Quantum Information (IQOQI), Austrian Academy of Sciences, Boltzmanngasse 3, 1090 Vienna, Austria}
\affiliation{Vienna Center for Quantum Science and Technology (VCQ), Faculty of Physics, University of Vienna, Boltzmanngasse 5, A-1090 Vienna, Austria}

\date{\today}

\begin{abstract}

In entanglement based quantum key distribution (QKD), generation of entangled photon pairs and the random choice of measurement basis can be implemented with passive optical devices. However, auxiliary random numbers are still necessary for conducting the privacy amplification procedure. 
In this paper, we solve this issue and propose a fully passive entanglement based QKD scheme in which certified random bits for privacy amplification are extracted from detection events of measurement-basis-mismatched entangled photon pairs. 
The privacy of these random bits can be certified by estimating the phase and bit error rate in the sifted key. 
The idea of extracting certified random numbers from measurement-basis-mismatched photons, can be extended to all practical QKD schemes.

\end{abstract}

\pacs{Valid PACS appear here}
\maketitle


\section{\label{sec:Intro}Introduction}

Quantum Key Distribution (QKD) facilitates distributing unconditionally secure {random} keys between distant parties. There are many different types of QKD systems and one might distinguish between active and passive systems. Typical active systems implement decoy state QKD protocols, employing sources of weak coherent laser pulses (WCP)~\cite{Yin2016,Kiyoshi2016decoy,Zhao2006,Peng2007,Rosenberg2007,Yuan2007decoy,Liu2010decoy,Wang2013decoy,Tang2014decoy,Tang2014decoy200km} and require true random numbers for actively modulating certain parameters of the pulses (e.g. intensity, polarization, phase, etc.) and the choice of measurement settings, according to the needs of the specific QKD protocol. Contrary, it has been emphasized~\cite{curty2010passivedecoy} that passive systems might be advantageous in scenarios, where operation at a high transmission rate is required. Moreover, passive systems do not require the use of auxiliary random numbers and optical modulators, and are thus immune against attacks through side-channels provided by imperfections of these devices. Additionally, for the very same reason they provide the potential for significantly reducing the overall costs and complexity for the implementation of QKD systems. This can be advantageous for example when deploying QKD system on a satellite.
Many groups have proposed schemes to implement passive decoy state QKD systems based on parametric down conversion (PDC)~\cite{Rarity1994,Adachi2007,Mauerer2007,Ma2008,Wang2007,Adachi2009,tan2011,wang2016} as well as on WCP~\cite{curty2010passivedecoy,Wang2007wcp,curty2010passive84,Li2014,Shan2014,curty2015} and both types of sources have recently already been implemented~\cite{Guan2015,sun2016pra,sun2014,Krapick2014,Zhang2012,Zhang2010}. 
Another scheme for passive QKD, which does not rely on the generation of decoy states, is based on entanglement and has been implemented in a wide range of systems with the most prominent utilizing the polarization or the time degree of freedom~\cite{Jennewein2000,Tittel2000,Resch2005,Peng2005,Marcikic2004,Ursin2007,Poppe2004,Maentangle2007,Marcikic2006,Ali2006,Honjo2008,Fasel2004,Alexander2009,Takesue2010,scheidl2009}. Even though passive QKD systems do not require auxiliary true random numbers for the state generation and also the random choice of the measurement basis can easily be implemented in a passive way using e.g. a $50/50$ beam splitter, true random numbers are still  {necessary} for conducting the most important post-processing procedure -- privacy amplification~\cite{Bennett1988, Bennett1995, Watanabe2007, boliu2013}. 
Privacy amplification reduces the mutual information between the eavesdropper and the communication parties~\cite{Bennett1988, Bennett1995, Fung2010Post, LIUrealtime2013, boliu2013, boliu2016, hayashi2011, Watanabe2007, Renner2005PA} by applying a randomly chosen universal hash function to the sifted and error-corrected key. Therefore, the universal hash function is usually constructed with the help of a {trusted} auxiliary random source~\cite{Hayashi2009,hayashi2011,hayashi2014,hayashi2015}. 

Since the generation of true random numbers must rely on an unpredictable physical process~\cite{Pironio2010}, great effort has been put into the investigation and development of quantum random number generators (QRNGs), based on the intrinsic randomness of the quantum measurement process~\cite{Jennewein2000,Fiorentino2007,Wei2009QRNG,Svozil2009QRNG,Furst2010,Jofre2011,Gallego2013QRNG,Abellan2014,Dhara2013,Pironio2010,Marangon2017,Acin2016Nature,Acin2016PRA,Vallone2014,Rarity1994,Herrero2017,2020arXiv200212295P}. 
Today, high quality devices are already commercially available passing the statistical tests on the randomness of the output bits. However, statistical tests alone can not certify that the numbers are not known to an adversary~\cite{Marangon2017}. 
A sequence of bits is thought to be ``certified random'', if it is perfectly random and unpredictable by any observer. In the device independent frame work, it has been shown that random numbers can be generated utilizing entangled states, while the privacy can simultaneously be certified via the violation of a Bell inequality~\cite{Acin2016Nature}. 
We want to point out that the device independent generation of certified random numbers essentially requires a loophole-free violation of a Bell inequality and is thus very hard to achieve in any practical system. 
Meanwhile, under the assumption of trusted or well-known and characterized measurement devices, the privacy of random numbers can be certified by performing given positive-operator-valued measures~\cite{Vallone2014,Acin2016PRA}.

Here we propose a fully passive entanglement-based QKD scheme in which certified random bits for privacy amplification are extracted from detection events of measurement-basis-mismatched photons. The randomness is guaranteed by the intrinsic randomness of the outcomes of measurements on entangled states while, building on the very same assumptions on which the security of any practical QKD system is based on, the privacy of these random bits can be certified by estimating the phase and bit error rate in the sifted key. 
The idea of extracting certified random numbers from measurement-basis-mismatched photons is very attractive for satellite-based QKD systems, which provides the potential for significantly simplify the implementation of QKD payload on satellite. 

\section{\label{Sec:Preliminaries}Preliminaries}

The security of practical QKD systems is based on the following underlying assumptions:

\begin{enumerate}
 
  \item The quantum formalism is valid and all devices as well as the adversary are constrained by the laws of quantum mechanics.
  \item Alice and Bob share a certain amount of secure key prior to starting the QKD session.
  \item Alice and Bob share independent and identically distributed (IID) quantum systems and know the dimension of the Hilbert space describing these systems~\cite{Antonio2006Bell}.
  \item The detection system is compatible with the squashing model~\cite{Beaudry2008,Gittsovich2014}.

\end{enumerate}

Usually, communication parties of practical QKD systems should also be equipped with trusted random number generators and trusted key management. For the practical entanglement based QKD scenario, Alice and Bob have to trust their local measurement systems. 

The standard implementation of the entanglement based QKD protocol mainly consists of four phases: \emph{quantum communication phase}, \emph{basis reconciliation}, \emph{error correction} and \emph{privacy amplification}. 
In the quantum communication phase, the communicating parties Alice and Bob share pairs of polarization entangled photons, which are passively generated at a separated source and distributed through quantum channels. 
Alice and Bob each measure every impinging photon randomly in one of two complementary polarization measurement bases and associate the two possible measurement outcomes in any measurement basis with the binary values ``0'' and ``1'', respectively. 
The random basis choice is either implemented with active polarization switches controlled by local random number generators or with passive devices like a 50/50 beam splitter. Finally, Alice and Bob only keep those results where both photons of a pair have been detected. This is called the raw key $K_r$ with length $n_r$.   

The quantum communication phase is followed by basis reconciliation, where Alice and Bob publicly exchange information about the measurement bases they actually implemented for their detected photons. 
Owing to the properties of the maximally entangled state, they {typically} discard all bits for which their measurement bases did not match, since in these cases their measurement results are maximally uncorrelated. Contrary, their measurement results are perfectly (anti-)correlated when they used the same bases. In doing so, Alice and Bob end up with the sifted key $K_{sift}$ with length $n_s$. 

In a realistic scenario, $K_{sift}$ always contains errors, which all have to be attributed to a potential eavesdropping attack. During error correction, these errors are corrected implementing classical algorithms~\cite{BBBSS1992,Brassard1994Cascade,Sugimoto2000Cascade,Yan2008Cascade,Pedersen2015Cascade, Buttler2003Winnow,CuiWinnow2013, Elkouss2009LDPC, Wilde2013Polar} via public channels, such that Alice and Bob end up sharing an identical key $K_{EC}$. Finally, all information an eavesdropper might have obtained during the communication phase or error correction is removed with the help of privacy amplification.

Privacy amplification is a procedure allowing two communication parties, which share a weak secret key $K_{EC}$ to agree on a final secrete key $K_f$ of length $n_f$ via communication over a authenticated public channel. Thereby, the public channel might even be controlled by the adversary Eve, who is granted unbounded computational power. Let us denote $E$ an arbitrarily string that Eve may have learned during the QKD procedure. During privacy amplification, Alice typically uses an auxiliary random number generator to generate a random bit string and sends it to Bob through a public channel. Then, Alice and Bob use this random string to construct a universal hash function, which they apply to $K_{EC}$. 

Even in QKD protocols with passive quantum communication phase, other sub-procedures of the protocol often require local randomness and therefore QKD implementations are usually equipped with proprietary random number generators. Consider for example, squashing operation for multiple-click events, which often require some local randomness, because double-click events in one basis are typically assigned a random outcome value. Another example is error correction where some protocols use local randomness as a free resource (e.g. CASCADE protocol introduced in \cite{Brassard1994Cascade}). Additionally, in determining the error rates of practical QKD protocols, Alice and Bob need to exchange small randomly chosen subset of their raw key. Last but not least, as mentioned above, privacy amplification step requires local randomness in order to construct a universal hash function. 

In order to present a fully passive protocol in this manuscript we consider sub-procedures, which do not require local randomness wherever possible. For example, one can count the number of double-click events and perform privacy amplification based on the counted number of the double-click events~\cite{Gittsovich2014,RN1884} without resorting to extra local random strings, or use error correction based on standardized LDPC codes, which also does not utilize additional randomness \cite{mink2014}. Further, since error rates can be well estimated from logarithmically small samples, local randomness needed to choose them can be neglected compared to privacy amplification. This leaves privacy amplification as the only sub-procedure which requires a significant amounts of local randomness. In this paper, we show that enough local randomness for privacy amplification can be obtained by utilising measurement outcomes in mismatched measurement bases of Alice and Bob, which are discarded in a typical protocol.

Let us denote the length of the final key $K_f$ as $n_f$. Then, $n_f$ is given by
\begin{equation}
\label{equ:keyrate}
n_f = n_f^x + n_f^z,
\end{equation}
{where $n_f^x$ ($n_f^z$) is the final secure key length extracted in $X$ ($Z$) basis, which can be obtained using~\cite{Maentangle2007}:
\begin{align}
n_f^x &\geq n_s^x\left[H_\text{min}^x(K_{sift}^A\vert E)-f\left(e_{bx}\right)H^x\left(K_{sift}^A\vert K_{sift}^B\right)\right]\nonumber\\\label{equ:keyratex}
&\geq n_s^x\left[1-H_2\left(e_{px}^{U}\right)-f\left(e_{bx}\right)H_2\left(e_{bx}\right)\right].
\end{align}
Here $H_\text{min}^x(K_{sift}^A| E)$ is min-entropy of a single Alice's sifted key bit obtained by measurement in the $X$ basis conditioned on the knowledge of the adversary and $H^x\left(K_{sift}^A\vert K_{sift}^B\right)$ is the error rate between sifted keys of Alice and Bob, obtained by measurement in the $X$ basis. Further, $n_s^x$ is the number of bits in the sifted key obatained by measurement in the $X$ basis, $e_{px}^U$ and $e_{bx}$  are the estimated upper bound of phase error rate and the measured quantum bit error rate (QBER) in $X$ basis, $f(\cdot)$ is the error correction efficiency and $H_2(\cdot)$ is the binary entropy function.
Analogously, the length of the part of the final key, which Alice and Bob can obtain by measurement in $Z$ basis is defined as~\cite{Maentangle2007}:
\begin{align}
n_f^z &\geq n_s^z\left[H_\text{min}^z(K_{sift}^A\vert E)-f\left(e_{bz}\right)H^z\left(K_{sift}^A\vert K_{sift}^B\right)\right]\nonumber\\
&\geq n_s^z\left[1-H_2\left(e_{pz}^{U}\right)-f\left(e_{bz}\right)H_2\left(e_{bz}\right)\right] \label{equ:keyratez}.
\end{align}}

\section{\label{Sec:Scheme}Passive Entanglement Based QKD Scheme}

\begin{figure}
    \includegraphics[width=1\columnwidth]{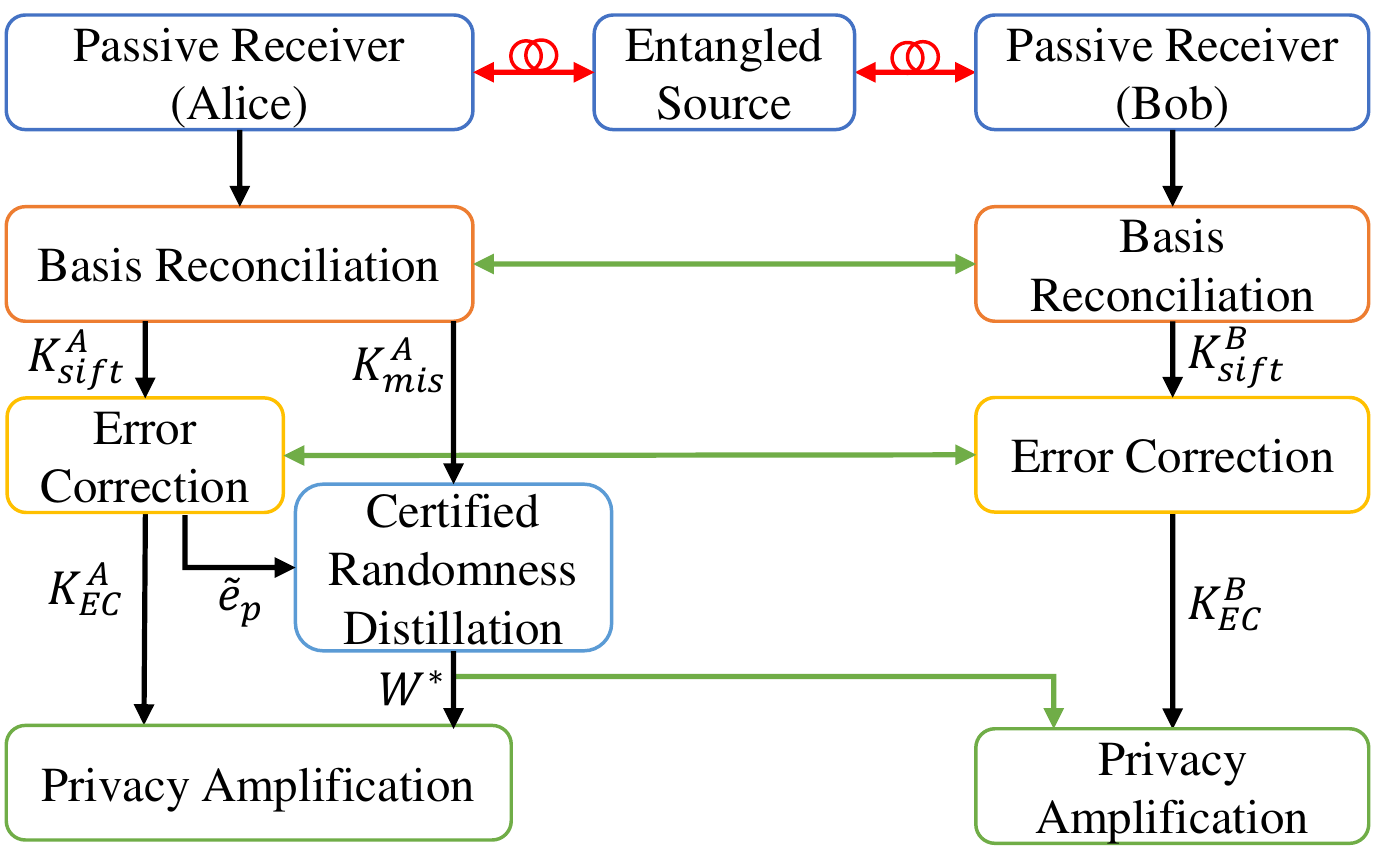} 
    \caption{Fully passive entanglement based quantum key distribution scheme. $K_{mis}^A$ is the reserved sifted key string, which is measured in mismatched basis between Alice and Bob. Random seed $W^\ast$ is used to publicly choose the universal hash function. $K_{sift}$ is the key string after the basis reconciliation. $K_{EC}$ is the key string after the error correction phase. $\widetilde{e}_p$ is the estimated phase error rate. Entangled photons are transmitted to Alice and Bob through the quantum channel, specified in red lines. Classical information exchange between Alice and Bob is by the ITS authenticated classical channel, specified in green lines.}
    \label{Fig:MEPA}
\end{figure}

The fully passive entanglement based QKD scheme is shown in Fig.~\ref{Fig:MEPA}. Entangled photon pairs can be produced by parametric down conversion with nonlinear crystals. With beam splitter, polarizing beam splitter and other passive optical elements, Alice and Bob implement the passive receiver for detecting the entangled photon pairs, in particular, the measurement basis choice is implemented with a $50/50$ beam splitter. Thus, the quantum physical communication between Alice and Bob is done in a fully passive way. Here, we focus on how to conduct privacy amplification with the certified random bits extracted from measurement-basis-mismatched photons, which mainly includes four steps.

{\bf Step 1. Generation of local random bit-strings.} First, during the basis reconciliation procedure, one has to reserve the bit string $K_{mis}$ at Alice's or Bob's side, respectively, resulting from local measurement results, which have been detected in measurement-basis-mismatched photons. In this paper, we gather Alice's mismatched basis measurement results into a string labeled $W=K_{mis}^A$.
Intuitively, this string contains as much randomness and privacy as the $K_{sift}$, which can be used in the privacy amplification part of the protocol.

{\bf Step 2. Randomness certification.} The amount of information about $W$, which is unknown to anybody else before the post-processing procedure started (under the assumptions in Sec.~\ref{Sec:Preliminaries}) is given by its min-entropy. Hence, the goal is to find a lower bound for $H_{\min}\left(W|E\right)$, which allows to certify its randomness and privacy.

As an adversary cannot influence the local measurement devices, her attack strategy is the same for photon pairs measured in either matched or mismatched bases. This means we have 
\begin{align}\nonumber
 \!H_{\min}\!\left(W|E\right)\! &=\! m_x H_{\min}^x\!\left(K_{sift}^A|E\right)  \!+\! m_z H_{\min}^z\!\left(K_{sift}^A|E\right)\\\label{equ:MinentropyKmisX}
 &=\!m_x\!\left[1\!-\!H_2\!\left(e_{pz}^{U}\right)\right]\!+\!m_z\! \left[1\!-\!H_2\!\left(e_{px}^{U}\right)\right]\!,
\end{align}
where $m_x$ and $m_z$ are the number of measurement results obtained in $X$ and $Z$ basis in the mismatched rounds. The second equation was obtained by using (\ref{equ:keyratex}) and (\ref{equ:keyratez}) and omitting the error correction cost $f\left(e_{bz}\right)H_2\left(e_{bz}\right)$ and $f\left(e_{bx}\right)H_2\left(e_{bx}\right)$, since the local randomness does not need error correction.

After estimating the min-entropy of $W$ from the phase error rates ${e_{px}^U}$ and ${e_{pz}^U}$, an adequate randomness extractor has to be implemented to distill perfect and uniformly distributed random string $W^\ast$ \cite{Herrero2017}. 
This can be done with the use of the seeded extractors~\cite{trevisan2001extractors,ma2013,Konig2011Ext,Skorski2015Ext} with the help of some additional randomness. This additional randomness can either be taken from the pre-shared secure key, or the protocol can start with a short local random string. {The main idea is that the seed used to extract uniform local randomness from $W$ is not made public, therefore it can be reused multiple times with help of leftover hashing lemma \cite{ma2013}. This is important, since it shows that the amount of local randomness this extraction step consumes is negligible.} 

{\bf Step 3. Public Hash Function Discussion.} Alice sends the extracted random seed $W^\ast$ to Bob via a authenticated public channel.

{\bf Step 4. Key Extraction.} Alice and Bob use  $W^\ast$  to construct a universal hash function for privacy amplification, which they apply to their error corrected key $K_{EC}$ resulting the final secure key $K_f$.

{In order to evaluate the key rate of our protocol, we need to specify the family of hash functions used in the privacy amplification. This choice crucially influences the amount of local uniform randomness $W^*$ needed to extract the final key.
The main idea is that if the mismatched measurements do not produce enough local randomness in order to extract $K_f$ from $K_{EC}$, Alice will reassign $\epsilon$ bits from $K_{EC}$ to $W$. By doing so she reduces the amount of local randomness needed for privacy amplification, which usually depends on the length of $K_{EC}$ and/or its min-entropy, both of which decrease by reassigning key bits to local randomness, while simultaneously increasing the amount of local randomness in $W$.
Let us define a hash-family-specific function $h(n_s,n_f)$, which returns the size of the local seed needed to extract the final key of length $n_f$ (which is equivalent to the total min-entropy of $K_{EC}$) from the error corrected key of length $n_s$. 
Then the length of $K^A_mis$ is $(n_r-n_s)$. 
Further, for simplicity let us denote  
$\widetilde{e}_p := \max(e_{pz}^U, e_{px}^U)$ and $\widetilde{e}_b:= \max(e_{bz}, e_{bx})$. Then with $K_{raw}$ of length $n_r$ and $K_{EC}$ of length $n_s$, we have
\begin{align}
    H_\text{min}(W|E) = (n_r - n_s)[1-H_2(\widetilde{e}_p)]\label{equ:minentropyAgregate}
\end{align}
and the total entropy of Alice's error corrected key is
\begin{align}
    H_\text{min}(K_{EC}^A|E) = n_s[1-H_2(\widetilde{e}_{p})-f(\widetilde{e}_b)H_2(\widetilde{e}_b)].\label{equ:keyRateAggregate}
\end{align}
Note that we use $\widetilde{e}_p$ and $\widetilde{e}_b$ only for convenience of notation, since it allows us to disregard the information about the measurement basis. The result can be generalized in a straightforward way to take into account also the information in which basis the measurement outcomes forming both $W$ and $K_{EC}^A$ were obtained.}
Taking into account the procedure to enlarge $W$ by $\epsilon$ bits, together with equations \eqref{equ:minentropyAgregate} and \eqref{equ:keyRateAggregate} the final secure key length of fully passive scheme is defined as
\begin{equation}
\label{equ:KeyRatepassive}
n_f^\ast \geq (n_s-\epsilon)[1-H_2(\widetilde{e}_{p})-f(\widetilde{e}_b)H_2(\widetilde{e}_b)],
\end{equation}
where $\epsilon$ is the smallest integer such that
\begin{align}
   {[n_r\!-\!n_s+\epsilon]}&{[1\!-\!H_2(\widetilde{e}_p)]}\!\geq\\\nonumber &{h\left[n_s\!-\!\epsilon,\!(n_s\!-\!\epsilon)[1\!-\!H_2(\widetilde{e}_{p})\!-\!f(\widetilde{e}_b)H_2(\widetilde{e}_b)]\right],}
\end{align}
i.e. the smallest $\epsilon$ that leads to a sufficient length of the local random string $W^*$ usable as a seed for hash function with requirements $h(n_s,n_f)$.
In order for this paper to be self sufficient, in table \ref{Tab:hash} we list a table of functions $h(n_s,n_f)$ for different families of hash functions studied in \cite{hayashi2014}. 
\begin{table}[tbh]
\caption{Seed requirements of different families of hash functions.}
\label{Tab:hash}
\begin{tabular}{ll}
\hline\hline
Hash function family & $h(n_s,n_f)$ \\\hline
$f_{F1,R}$ and $f_{F2,R}$ \cite{hayashi2014}& $n_s-n_f$ \\
$f_{F3,R}$ and $f_{F4,R}$ \cite{hayashi2014} & $n_f$ \\
Hash functions using Toeplitz matrix & $n_s$ \\
Trevisian's extractor~\cite{RN1015} & $\log^3(n_s)$ \\
Hash functions in the TSSR paper~\cite{RN655} & $2n_f$ \\
$\epsilon$-almost pairwise independent hash functions~\cite{RN863} & $4n_f$\\
\hline\hline
\end{tabular}
\end{table}

\section{Simulation Results}

We simulate the performance of the fully passive entanglement based QKD with the parameters shown in Table.~\ref{Tab:para}. We compare fully passive and standard entanglement (BBM92) QKD schemes~\cite{Bennett1992}. In order to establish optimal secure key rate, we take the optimal photon pair number per coincidence window ($\mu$). Finite-size-effect is considered with post-processing block size equals to $10^6$ and the failure probability $\varepsilon^{ph}=10^{-7}$ for estimating the phase error rates. 

\begin{table}[tbh]
\caption{Parameters used for simulation.}
\label{Tab:para}
\begin{tabular}{ll}
\hline\hline
Parameters & Values \\\hline
Dark Count Rate $p_d$ & $10^{-6}$ \\
Detector Efficiency $\eta_d$ & 0.40 \\
Misalignment error rate $e_d$ & 0.015 \\
Error Correction Efficiency $f$ & 1.15 \\
Photon Pair Number per Coincidence Window $\mu$ & Optimal \\
Basis Reconciliation Factor $q$ & 0.50\\
Phase Error Estimation Failure Probability $\varepsilon^{ph}$ & $10^{-7}$\\
Post-processing Block Size & $10^{6}$\\
\hline\hline
\end{tabular}
\end{table}

\begin{figure}[tb]
    \includegraphics[width=1\columnwidth]{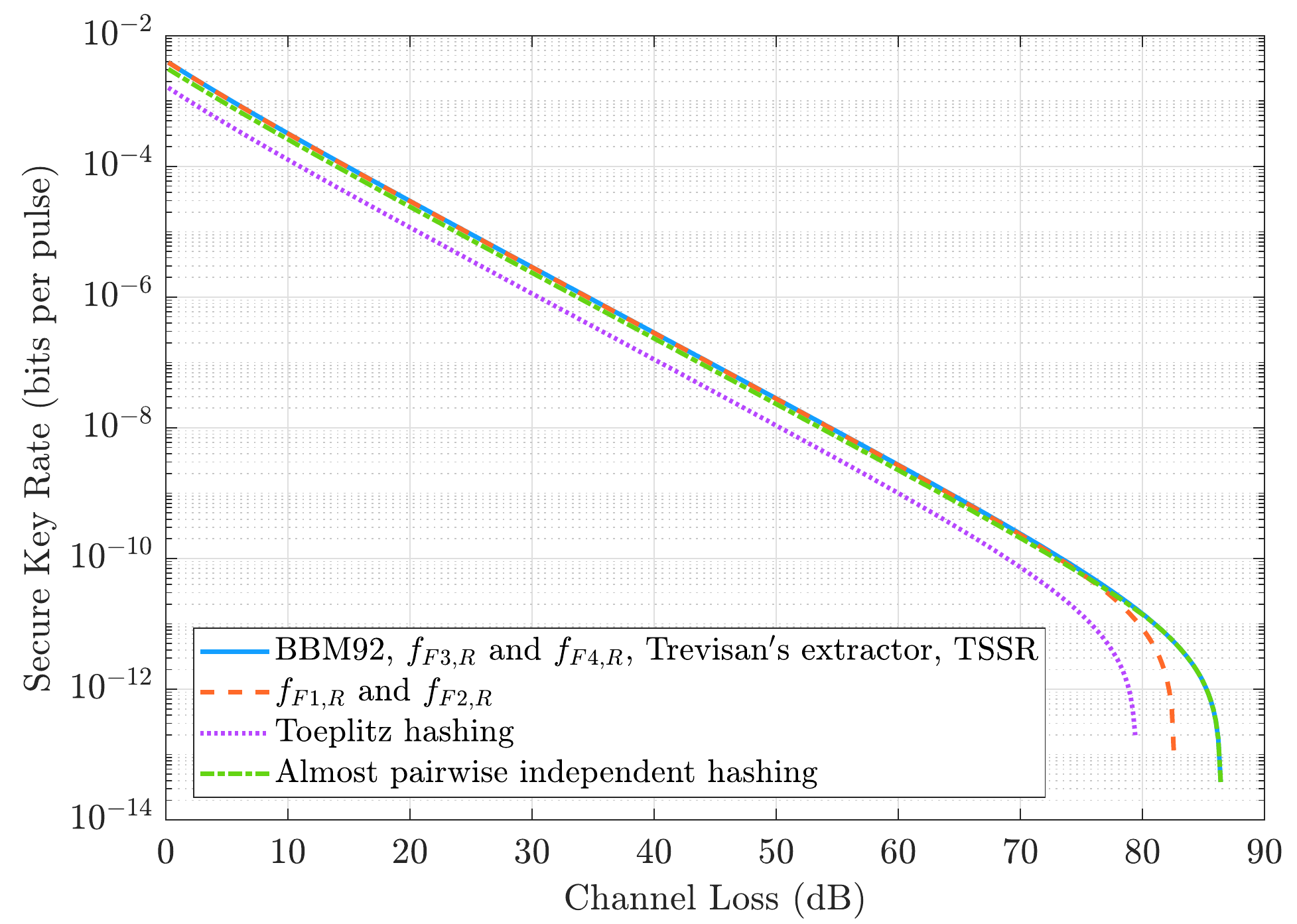} 
    \caption{a) Secure key rate per pulse versus channel loss (source in between). For each secure key rate plot, we numerically optimize the mean photon pair number $\mu$. The hash function $h$ performed in this paper is shown in Table~\ref{Tab:hash}.}
    \label{Fig:R_in_Between}
\end{figure}

The simulation results are shown in Fig.~\ref{Fig:R_in_Between}. The secure key rate per pulse of fully passive entanglement based QKD scheme is varying a lot depending on the performed hash function in the privacy amplification procedure. When $f_{F3,R}$, $f_{F4,R}$, TSSR hash functions or Trevisan's extractor are used, the fully passive entanglement based QKD scheme achieves the same secure key extracting rate as the BBM92 scheme. Considering the computational complexity of the privacy amplification procedure, we suggest the implementation scheme with hash functions in family $f_{F3,R}$ and $f_{F4,R}$, which achieve computational complexity of $O(n\log n)$.

\section{Discussion}

We described a fully passive entanglement based QKD protocol, which does not require any auxiliary random number generators. We achieve this by producing certified local random bits extracted from mismatched basis measurements during a run of the protocol. These measurement-basis-mismatched events are typically discarded during the basis reconciliation procedure of QKD protocols. However, as we show above, these events can be used to produce a local random string, which can afterwards, be used in  privacy amplification procedure. To our knowledge, this simple but powerful observation has never been mentioned in literature before. The advantages of a fully passive protocol are twofold. Firstly, since auxiliary random number generators and active optical modulators are not utilized, our protocol is robust against attacks through side-channels caused by imperfections of these devices.
Secondly, the fact that our protocol does not require auxiliary random numbers provides the potential for significantly reducing the overall costs and complexity for the implementation of QKD systems, which can be beneficial for example in reducing payload on a satellite. 
Last but not least, the idea of utilizing mismatched measurement results is very simple and can be used in conjunction with plethora of existing practical QKD protocols.

\begin{acknowledgments}

We thank Anton Zeilinger for for fruitful discussions. We acknowledge funding from the FWF START project (Y879-N27) and the FWF-GACR joint international project (I3053-N27 and GF17-33780L), the Austrian Research Promotion Agency (FFG) via the Austrian Science and Applications Programme (ASAP12, Contract 6238191), the Austrian Federal Ministry of Education, Science and Research (BMBWF) and the University of Vienna via the project QUESS, the National Natural Science Foundation of China via project No. 61972410 as well as from the research plan of National University of Defense Technology under Grant No. ZK19-13 and No. 19-QNCXJ-107.

\end{acknowledgments}

%

\end{document}